**Kinetic and H-atom product study of the C + $C_6H_6$ reaction at low temperatures.**

Kevin M. Hickson[1,*] and Jean-Christophe Loison[1]

[1] Univ. Bordeaux, CNRS, Bordeaux INP, ISM, UMR 5255, F-33400 Talence, France

*Email: kevin.hickson@u-bordeaux.fr

**Abstract**

The reactions of atomic carbon in its ground electronic state configuration, $C(^3P)$, are potentially important processes in astrochemistry due to the large abundance of $C(^3P)$ atoms in the interstellar medium (ISM) and its high overall reactivity towards a wide range of molecules. Although benzene, $C_6H_6$, has not been detected in the dense ISM, its presence at high abundance levels is inferred through the detection of functionalized derivatives. Here we present a combined experimental and astrochemical modeling investigation of the gas-phase $C(^3P) + C_6H_6$ reaction. Experimentally, rate constants were determined over the 50-296 K range using the Laval nozzle technique coupled with pulsed laser photolysis and laser induced fluorescence for $C(^3P)$ generation and detection respectively. Product yields of atomic hydrogen, $H(^2S)$ were also measured at 177 and 296 K to provide some information on the product channels of the reaction. The measured rate constants are very large, between 3.3 and $5.7 \times 10^{-10}$ $cm^3$ $s^{-1}$ indicating the fast, barrierless nature of the reaction, while the H-atom yields are all below 10 % when compared to the $C(^3P) + C_2H_4$ reference reaction. As the $C(^3P) + C_6H_6$ reaction is not currently included in astrochemical databases, its influence on the simulated abundances of $C_6H_6$ and related species was tested using a gas-grain model of dense interstellar clouds. The $C(^3P) + C_6H_6$ reaction is shown to be the main loss process for interstellar benzene, while different hypotheses regarding the product channels are discussed in the context of observations to shed light on the preferred formation pathways.

## 1 Introduction

Polycyclic aromatic hydrocarbons (PAHs) are widely considered to be important molecules in the chemistry of the interstellar medium and planetary atmospheres. It has been suggested that PAHs might be responsible for the various absorption bands stretching from the near ultraviolet to the near-infrared region known as the diffuse interstellar bands (DIBs), the unidentified infrared emission bands (UIRs) and the aromatic infrared bands (AIBs).[1] They could also contain a significant percentage of the overall carbon budget of the interstellar medium (10-20%).[1] Chemical models simulating the formation of species in interstellar environments through a bottom-up approach generally consider that benzene, $C_6H_6$, is an essential precursor for generating larger PAHs through its various low temperature reactions.[2] $C_6H_6$ has been positively identified in circumstellar media in the direction of the proto-planetary nebula CRL 618 by infrared spectroscopy using the Infrared Space Observatory,[3] in the atmosphere of Saturn's moon Titan, [4, 5] and in the disk surrounding a very low mass protostar.[6] Although $C_6H_6$ has never been identified in cold interstellar medium (ISM) due to the absence of a permanent dipole moment, making it undetectable by rotational spectroscopy, its presence has been inferred by the detection of functionalized benzene derivatives such as cyanobenzene ($C_6H_5CN$)[7] which possesses a large dipole moment. Previous work has clearly demonstrated that the reaction between the CN radical and $C_6H_6$ is rapid at low temperature[8, 9] leading to $C_6H_5CN$ formation.[2] McGuire et al.[7] underpredicted the observed abundance of $C_6H_5CN$ of $1.65 \times 10^{-10}$ (abundances are relative to $H_2$) in TMC-1[10] using the gas-grain chemical model Nautilus,[11] based on parameters from the KIDA database.[12] These simulations predicted that $C_6H_6$ should reach a peak abundance of around $7 \times 10^{-10}$. Additional modifications to the chemical network allowed the agreement between the model and the observations to be improved further.[13] Recently, Loison et al.[14] revisited the chemical network for the production of benzene and its derivatives, simulating $C_6H_5CN$ and $C_6H_5C_2H$ with an improved factor of better than 2. The calculated abundance of benzene reached $2 \times 10^{-9}$. In order to evaluate the relevance of $C_6H_6$ to the formation of larger PAHs and other complex organic molecules, and to place more stringent constraints on their formation pathways, it is important to study the reactions of $C_6H_6$ with a range of abundant interstellar species. In particular, the reactions of $C_6H_6$ with carbon bearing radicals could be important sources of large aromatic and aliphatic organic molecules, and such reactions have been the focus of several previous studies.

Becker et al.[15] studied the kinetics of the reaction between dicarbon in its first excited triplet state, $^3C_2$ with benzene at room temperature, measuring a rate constant of $(4.9 \pm 0.1) \times 10^{-10}$ cm$^3$ s$^{-1}$. Berman et al.[16] measured rate constants for the reaction between the methylidyne radical, CH, and benzene using a slow-flow reaction vessel over the temperature range 297-674 K. They obtained a large rate constant for this process of $(4.3 \pm 0.3) \times 10^{-10}$ cm$^3$ s$^{-1}$, which was independent of temperature. Hamon et al.[17] measured the rate constant for the CH + $C_6H_6$ reaction at 25 K using a Laval nozzle reactor. They obtained a value of $(2.7 \pm 0.4) \times 10^{-10}$ cm$^3$ s$^{-1}$. Goulay et al.[18] studied the kinetics of the ethynyl radical $C_2H$ reaction with benzene between 298 and 105 K using a pulsed Laval nozzle reactor. They measured a fast rate constant for this process with a very slight negative temperature dependence with values in the range (3.3-4.2) $\times 10^{-10}$ cm$^3$ s$^{-1}$. Trevitt et al.[9] performed a detailed investigation of the CN + $C_6H_6$ reaction, measuring both temperature dependent rate constants for this process over the 105-296 K range using a pulsed Laval nozzle reactor and product channels at room temperature using synchrotron VUV photoionization mass spectrometry. In common with other studies of $C_6H_6$ reactivity, they derived large rate constants for this reaction with values mostly independent of temperature in the range (3.9-4.9) × $10^{-10}$ cm$^3$ s$^{-1}$ with the exclusive products $C_6H_5CN$ + H as described above. A more recent study of the CN + $C_6H_6$ reaction by Cooke et al.[8] over the 15-296 K range confirmed the rapid nature of the reaction down to interstellar temperatures with little or no temperature dependence. Atoms such H, N, C and O are among the most abundant of all species present in the ISM, with modeled abundances in the $10^{-4}$ – $10^{-6}$ range. While H, N and O have all been shown to be unreactive towards $C_6H_6$ at low and intermediate temperatures[19] and at low pressure, previous studies of the C + $C_6H_6$ reaction have demonstrated the potential importance of this process for the chemistry of the ISM. Haider and Husain[20] performed a room temperature kinetic study of this reaction using a flash photolysis apparatus coupled with atomic resonance absorption spectroscopy, detecting the decay of carbon atoms in the presence of an excess of $C_6H_6$. They measured a very fast rate constant of $(4.8 \pm 0.3) \times 10^{-10}$ cm$^3$ s$^{-1}$ for this process. Wenzel et al.[21] adopted this value of the rate constant in their astrochemical modeling study of cyanopyrenes. Kaiser et al.[22] studied the reaction of ground state atomic carbon, C($^3$P) with deuterated benzene $C_6D_6$ using a crossed molecular beam apparatus at an average collision energy of 32.1 kJ/mol. They observed a reactive scattering signal at m/z = 94 indicating the formation of $C_7D_5$ + D as products. These findings were supported by quantum chemical calculations indicating the formation of a $C_7D_6$ intermediate following C-attack on the $\pi$-electron density of the $C_6D_6$ molecule. A subsequent investigation of the C + $C_6H_6$ reaction at several different collision energies confirmed these preliminary findings and indicated the likely barrierless nature of the reaction over the triplet potential energy surface through additional electronic structure calculations.[23] A later study of the C + $C_6D_6$ reaction on liquid helium droplets[24] supported the conclusions of the earlier work of Kaiser et al.[22] by demonstrating the formation of the seven membered ring intermediate species $C_7D_6$. Bergeat and Loison[25] performed an experimental kinetic study of the C + $C_6H_6$ reaction at room temperature in a low-pressure fast-flow reactor and also derived relative atomic hydrogen yields by resonance fluorescence detection of $H(^2S)$ atoms at 121.567 nm. They measured a fast rate constant of $(2.8 \pm 0.8) \times 10^{-10}$ $cm^3$ $s^{-1}$ for this reaction somewhat lower than the value derived by Haider and Husain,[20] and a relative atomic hydrogen yield of $0.16 \pm 0.06$. These authors suggested that the low H-atom yield measured in their work (and apparent discrepancy with regard to the work of Kaiser et al.[22]) might originate from alternative product channels such as through the formation of c-$C_5H_4$ + $C_2H_2$ and/or intersystem crossing to the singlet surface to form the products $C_7H_4$ + $H_2$.

A computational study of the various pathways of the C + $C_6H_6$ reaction by da Silva[26] confirmed the existence of several exothermic H-atom elimination channels over the triplet potential energy surface (with the corresponding formation of several different $C_7H_5$ isomers) and also clearly identified the c-$C_5H_4$ + $C_2H_2$ exit channel with an energy of -52 kJ/mol with respect to the reagents based on geometries optimized at the M06-2X/6-31G(2df,p) level of theory with energies calculated using the composite G3X-K quantum chemical method. H-atom elimination channels over the triplet potential energy surface have also been studied by Izadi et al. 2021[27] showing that several isomers of $C_7H_5$ could be produced with similar branching ratios obtained from RRKM calculations.

Surprisingly, there are no previous measurements of the rate constants and product yields of the C + $C_6H_6$ reaction below room temperature. To address this issue, we report here the results of an experimental kinetic study of this reaction over the 296-50 K range to quantify its possible influence on interstellar chemistry. To perform these measurements, a continuous supersonic flow reactor was used, coupled with pulsed laser photolysis and pulsed laser induced fluorescence to generate C-atoms in-situ in the supersonic flow and detect them as a function of reaction time. In addition to the rate constant measurements, relative H-atom yields were also measured at two different temperatures to improve our understanding of the product channels of the C + $C_6H_6$ reaction. The new rate constants were then included in an astrochemical model to test their influence on the abundances of various interstellar species through various different scenarios.

The experimental methods used in this work are described in section 2, while the results of the kinetic and product studies are presented in section 3. The astrochemical simulations and the implications of this study for interstellar benzene and related species are given in section 4, followed by our conclusions in section 5.

## 2 Experimental Methods

The measurements presented here were performed on an existing continuous supersonic flow reactor described in detail in previous work, [28, 29] employing axisymmetric Laval nozzles as a means to create low temperature, high density ($10^{16}$ – $10^{17}$ $cm^{-3}$) flows so that thermal equilibrium is maintained. This system eliminates the condensation problems usually encountered by cryogenically cooled reaction vessels at low temperatures where condensable reagents are removed from the gas-phase through collisions with the cold reactor walls. The Laval nozzles are carefully designed to produce downstream flows of an inert carrier gas (Ar or $N_2$) with uniform temperature, density and velocity profiles as a function of distance from the nozzle exit for at least a few tens of centimetres, corresponding to useful hydrodynamic times of greater than 250 microseconds. Reagents and precursor molecules are introduced into these inert gas flows at low enough concentrations to avoid flow perturbations. Three Laval nozzles were used during the course of the present study, to produce 4 different low temperature flows of 177, 127, 75 and 50 K (one of the nozzles was used with both carrier gases). The properties of the flows generated by these nozzles are summarized in Table 1 of Hickson and Loison.[30] Experiments were also performed at room temperature by using the reaction chamber as a slow-flow reactor, without a Laval nozzle in the system.

Both the reagent molecules and precursor molecules were added to the reactor upstream of the Laval nozzle. $C_6H_6$ was introduced into the carrier gas flow through a bubbler system attached upstream of the Laval nozzle reservoir. Here a small calibrated flow of carrier gas was diverted into a bubbler containing $C_6H_6$ at room temperature. Then, the output flow carrying $C_6H_6$ vapour was passed into a cold trap to ensure the carrier gas was saturated at the cold trap temperature (290 K). The gas pressure in the bubbler and cold trap was carefully regulated around 200 Torr using a needle valve placed at the exit of the cold trap. The exit of the cold trap was connected to the reactor through a heated tube held at 353 K to prevent $C_6H_6$ condensation. The $C_6H_6$ vapour was diluted further upon reaching the reservoir, so we assume negligible condensation losses.

Carbon atoms were created in-situ within the cold supersonic flow for the experiments described here, with carbon tetrabromide, $CBr_4$, being used as the precursor molecule. The fourth harmonic of a pulsed (10 Hz) Nd:YAG laser at 266 nm was aligned along the length of the supersonic flow to create a column of C-atoms, through the pulsed multiphoton dissociation of $CBr_4$. Pulse energies of 35 mJ were typically used for this purpose with a 5mm diameter beam. $CBr_4$ was introduced into the system by flowing a small fraction (40-70 sccm) of the main carrier gas flow into a flask containing $CBr_4$ crystals. This flask was held at a known fixed pressure and room temperature for any single set of measurements, ensuring that the $CBr_4$ concentration in the supersonic flow was constant with a maximum value of $2.6 \times 10^{13}$ $cm^{-3}$ based on its saturated vapour pressure. Previous work[31] has shown that the 266 nm photodissociation of $CBr_4$ leads to the formation of atomic carbon in both the ground triplet state $C(^3P)$ and the first excited state $C(^1D)$, with $C(^1D)$ present at the 10-15 % level with respect to $C(^3P)$. The effect of the presence of $C(^1D)$ atoms in the cold supersonic flow on the kinetic measurements is negligible as discussed in earlier work,[32] while their effect on the H-atom product yield measurements is discussed in section 3.2.

Reagent $C(^3P)$ atoms in addition to product $H(^2S)$ atoms were detected by pulsed laser induced fluorescence in the vacuum ultraviolet wavelength range (VUV LIF) during the course of the present investigation. As the minor reagent species, $C(^3P)$ atoms were followed by exciting the $2s^22p^2\ ^3P_2 \rightarrow 2s^2\ 2p5d\ ^3D_3^\circ$ transition at 115.803 nm, with on-resonance detection of the fluorescence emission. Similarly, $H(^2S)$ product atoms were followed by VUV LIF through the Lyman-$\alpha$ transition at 121.567 nm. A multistep procedure was used to generate tunable radiation around these wavelengths. The output of a Nd:YAG pumped (excitation wavelength 532 nm) narrowband dye laser at 694.8 nm (729.4 nm for $H(^2S)$ detection) was steered into a beta barium borate crystal for second harmonic generation, producing radiation around 347.4 nm (364.7 nm for $H(^2S)$ detection). The beam exiting the crystal, containing both fundamental and second harmonic frequencies, was then reflected by two dichroic mirrors coated for maximum reflection around 355 nm allowing the fundamental dye laser radiation to be discarded. A small fraction of the discarded beam was fed into a fibre optic cable coupled to a wavemeter to allow for continuous precise monitoring of the excitation wavelength. The remaining UV beam was directed towards the reaction chamber by right angled prisms. Before entering the reaction chamber, the UV beam was first focused by a plano-convex lens (150 mm focal length) into a cell attached to the observation axis containing 50 Torr of xenon and 160 Torr of argon for frequency tripling purposes (210 Torr of krypton and 540 Torr of argon for H($^2$S) detection). A 75 cm long sidearm containing circular baffles was placed between the cell and the reactor. As the VUV radiation was divergent following its generation, a $MgF_2$ lens (50 mm focal length) was used as the exit window of the cell to collimate it. As a result of the different refractive indices for $MgF_2$ at UV and VUV wavelengths, the residual UV beam remained divergent after exiting the cell. The large distance between the cell and the reactor coupled with the use of baffles along the length of the sidearm meant that the flux of UV radiation reaching the reaction chamber was low. In order to avoid secondary absorption losses of the VUV excitation beam, a flush flow of Ar or $N_2$ was used to prevent reactive gases within the reactor from diffusing into the sidearm.

The VUV beam crossed the supersonic flow at right angles exciting the C($^3$P) or H($^2$S) atoms present in the illuminated volume. Fluorescence emission was collected by a solar blind photomultiplier tube (PMT) placed at right angles to the plane containing the VUV beam and the supersonic flow to minimize the detection of scattered light. A lithium fluoride (LiF) lens was placed in between the PMT and the supersonic flow to focus the fluorescence emission onto the PMT photocathode. The PMT itself was isolated from the chamber by a LiF window, thereby preventing degradation of the PMT entrance window, while the volume between the PMT and the LiF window was evacuated by an oil-free vacuum pump. A 500 MHz oscilloscope was used to monitor the PMT output signal and allow precise gate positioning for the boxcar integration system that was used to process it. The acquisition electronics and pulsed lasers were synchronized by a digital delay generator operating at 10 Hz, allowing the atomic fluorescence signals to be measured as a function of delay time between the photolysis and probe lasers. Each time point consisted of an average over thirty laser shots with at least 100 time points recorded to establish each temporal profile. The baseline level for the temporal profiles, consisting essentially of contributions from dark noise and scattered light, was set by firing the probe laser at several negative time delays with respect to the photolysis laser.

Carrier gases Ar (Messer 99.999 %) and $N_2$ (Messer 99.999 %) were regulated by calibrated mass-flow controllers. Rare gases Xe (Linde 99.999 %) and Kr (Linde 99.999 %) used for frequency tripling purposes, were taken directly from cylinders without further purification. Similarly, $C_6H_6$ (>99 %) was used directly from the bottle without further purification.

## 3 Results

### 3.1 Rate constants

All of the kinetic measurements described here were performed by applying the isolation method, by using a large excess concentration of $C_6H_6$ relative to atomic carbon. Under these conditions, the carbon atom concentration, which is considered to be proportional to its VUV LIF signal, decays exponentially as a function of reaction time, obeying the following first-order rate equation

$$I_C(t) = I_C \exp(-k_{1st}t) \qquad \text{E1}$$

where $I_C(t)$ and $I_C$ are the time dependent and initial C-atom fluorescence signals respectively, $k_{1st}$ is the pseudo-first-order rate constant for C-atom loss and $t$ is the reaction time corresponding to the time delay between photolysis and probe lasers. A couple of typical C-atom kinetic profiles, recorded at 50 K, are shown in Figure 1.

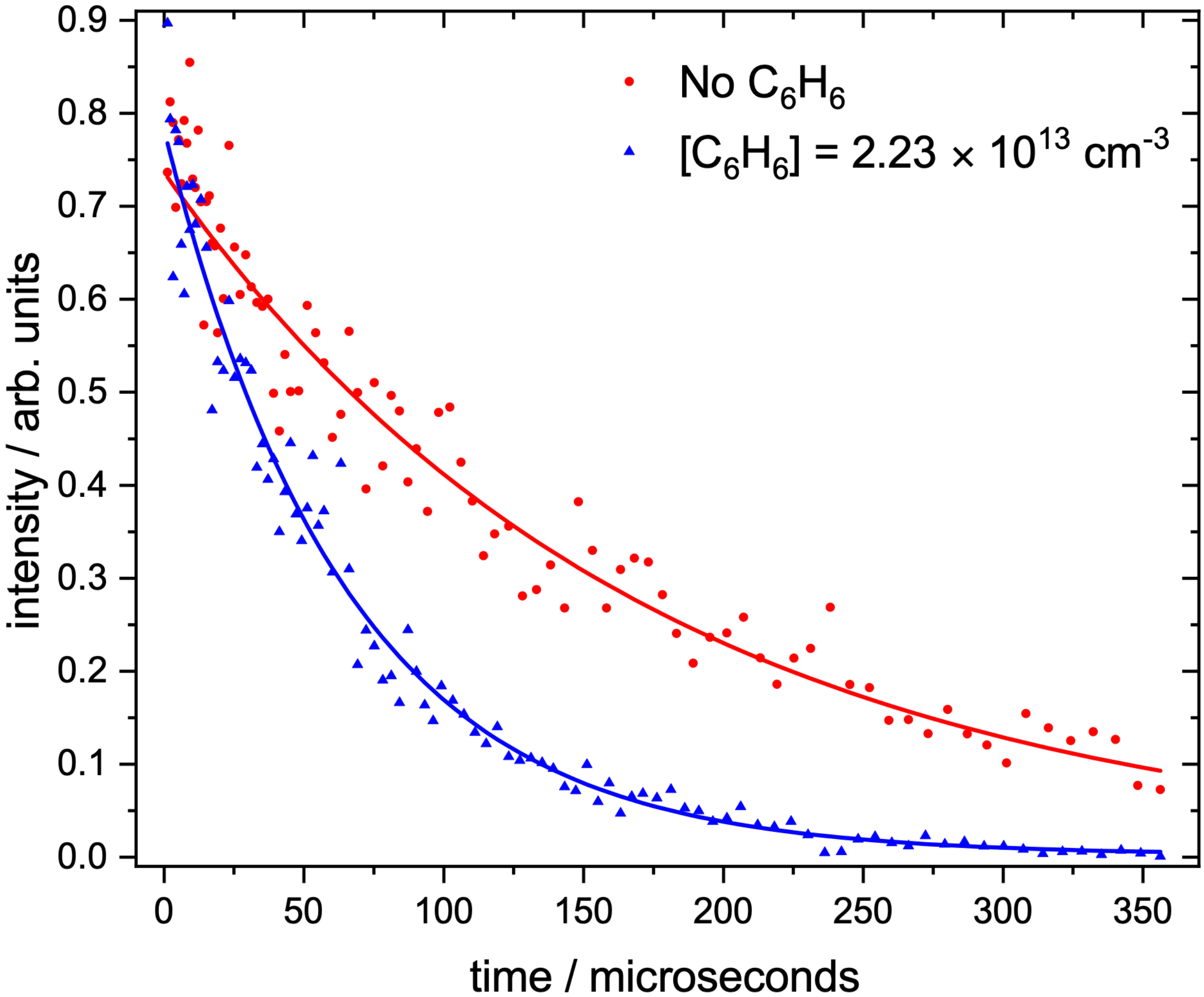

**Figure 1** C-atom VUV LIF signal as a function of time for two measurements performed at 50 K. (Red circles) C-atom decay in the absence of $C_6H_6$. (Blue triangles) C-atom decay with $[C_6H_6]$ = 2.23 × $10^{13}$ $cm^{-3}$. Solid red and blue lines represent exponential fits to the fluorescence profiles according to expression E1.

A non-linear least-squares fit to the temporal profiles using expression E1 allows the pseudo-first-order rate constant of the decay, $k_{1st}$, to be derived. Decay profiles were recorded for several different values of $[C_6H_6]$ and in the absence of $C_6H_6$. In addition, three separate temporal profiles were acquired for each $[C_6H_6]$. This procedure was repeated at all temperatures to obtain information on the temperature dependence of the reaction. In reality, there are several processes that contribute to rate constant for the loss of atomic carbon including:

$$k_{\mathrm{1st}} = k_{\mathrm{C+C_6H_6}}[\mathrm{C_6H_6}] + k_{\mathrm{C+CBr_4}}[\mathrm{CBr_4}] + k_{\mathrm{diff}} \qquad \text{E2}$$

where $k_{\mathrm{C+C_6H_6}}$ and $k_{\mathrm{C+CBr_4}}$ are the rate constants for the C + $C_6H_6$ and C + $CBr_4$ reactions respectively, and $k_{\mathrm{diff}}$ represents the diffusional loss of atomic carbon from the observation zone. For any single set of measurements, [$CBr_4$] is maintained at a fixed value, while $k_{\mathrm{diff}}$ is constant. Consequently, these terms have a fixed contribution to the overall decay constant. When the derived values of $k_{1st}$ are plotted as a function of [$C_6H_6$], as shown in Figure 2, it can be seen that the values of $k_{1st}$ vary linearly as a function of [$C_6H_6$] with the exception of the data recorded at 50 K for [$C_6H_6$] > $2.2 \times 10^{13}$ cm$^{-3}$.

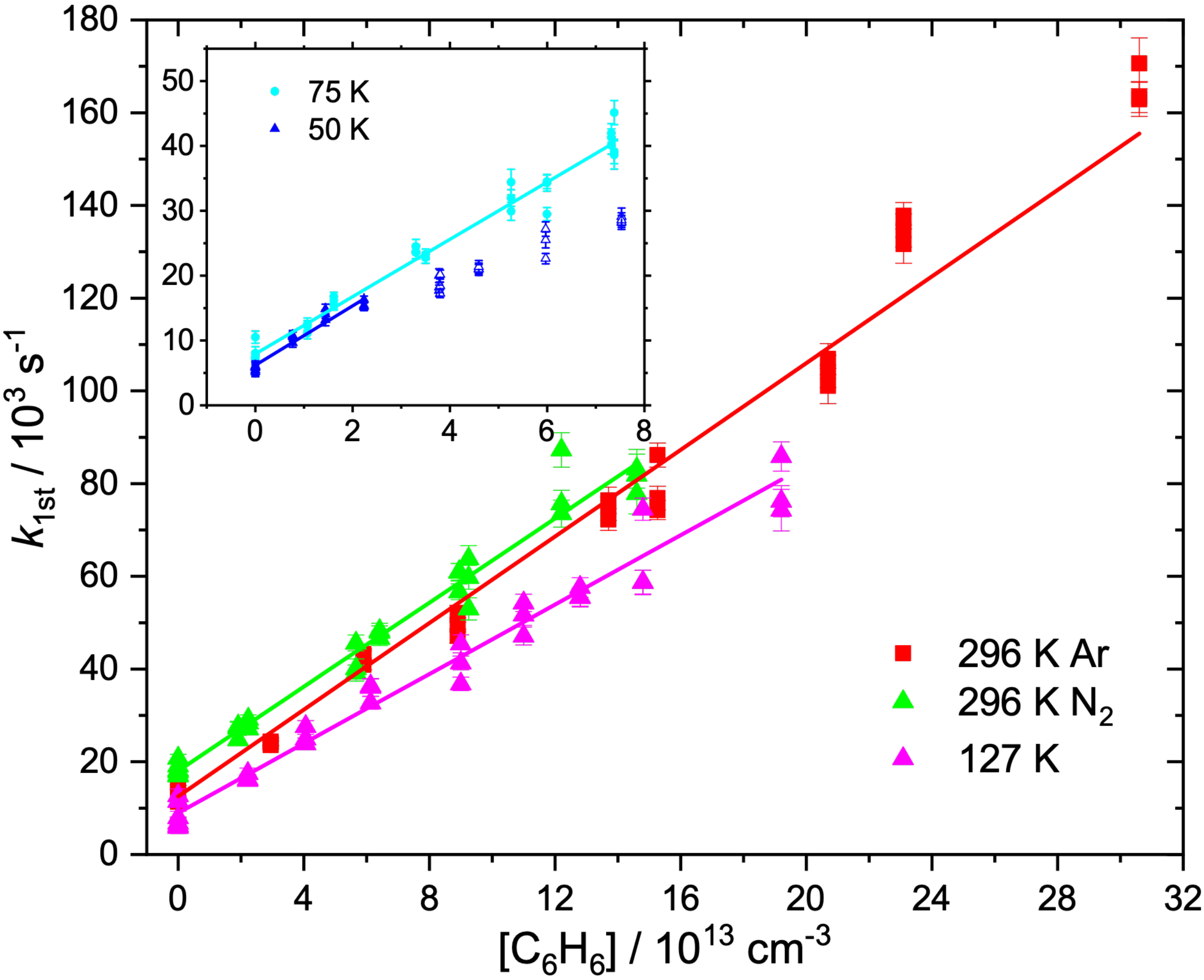

**Figure 2** Derived pseudo-first-order rate constants as a function of [$C_6H_6$] for a range of temperatures all measurements shown here are conducted with argon as the carrier gas unless mentioned otherwise. (Red solid squares) 296 K; (green solid triangles) 296 K, nitrogen carrier gas; (purple solid triangles) 127 K. Inset – expanded view of the lowest temperature data. (cyan solid circles) 75 K; (blue solid triangles) 50 K; (blue open triangles) 50 K data not used in the fit. Solid lines show the weighted linear-least squares fits to the individual data sets, with the gradient of these fits yielding the second-order rate constant. Error bars

represent the uncertainties associated with the pseudo-first-order rate constants derived from exponential fits to the atomic carbon fluorescence profiles such as those shown in Figure 1.

Here, the derived $k_{1st}$ values are significantly lower than expected from the linear fit to the $k_{1st}$ values obtained at [$C_6H_6$] < 2.2 × $10^{13}$ $cm^{-3}$, indicating the likely formation of $C_6H_6$ clusters in the supersonic flow, lowering the effective concentration of $C_6H_6$ monomers. Consequently, only $k_{1st}$ values obtained in the lower benzene concentration range were used to extract the second-order rate constant at this temperature. The temperature-dependent second-order rate constants were derived by a weighted linear least-squares fit to these data sets, with the statistical error σ determined during non-linear fits to the temporal profiles such as those shown in Figure 1 used as the weight according to the formula $1/\sigma^2$. The y-axis intercept value of these fits thus represents the sum of the second and third (constant) terms given by E2. The second-order rate constants measured during this work are summarized in Table 1 and these values are plotted as a function of temperature in Figure 3 alongside the rate constants derived in earlier work.

**Table 1** Measured second-order rate constants for the C + $C_6H_6$ reaction

| T / K | $N^b$ | [$C_6H_6$] / $10^{13}$ $cm^{-3}$ | Flow density] / $10^{17}$ $cm^{-3}$ | $k_{C+C_6H_6}$ / $10^{-10}$ $cm^3$ $s^{-1}$ | Carrier gas |
|---|---|---|---|---|---|
| 296 | 30 | 0 - 30.6 | 1.63 | (4.71 ± 0.49)[c] | Ar |
| 296 | 30 | 0 - 27.0 | 0.67 | (5.67 ± 0.58)[c] | Ar |
| 296 | 30 | 0 - 14.6 | 1.63,0.66[d] | (4.57 ± 0.47)[c] | $N_2$ |
| 177 ± 2[a] | 30 | 0 - 12.4 | 0.94 | (3.28 ± 0.34) | $N_2$ |
| 127 ± 2 | 30 | 0 - 19.2 | 1.26 | (3.98 ± 0.42) | Ar |
| 75 ± 2 | 30 | 0 - 7.4 | 1.47 | (4.41 ± 0.46) | Ar |
| 50 ± 1 | 15 | 0 - 2.2 | 2.59 | (4.79 ± 0.55) | Ar |

[a]Uncertainties on the calculated temperatures represent the statistical (1σ) errors obtained from Pitot tube measurements of the impact pressure.[29] [b]Number of individual measurements. [c]Uncertainties on the measured rate constants represent the combined statistical (1σ) and estimated systematic errors (10%). [d]Measurements performed at two different $N_2$ flow densities were combined to obtain the second-order rate constant.

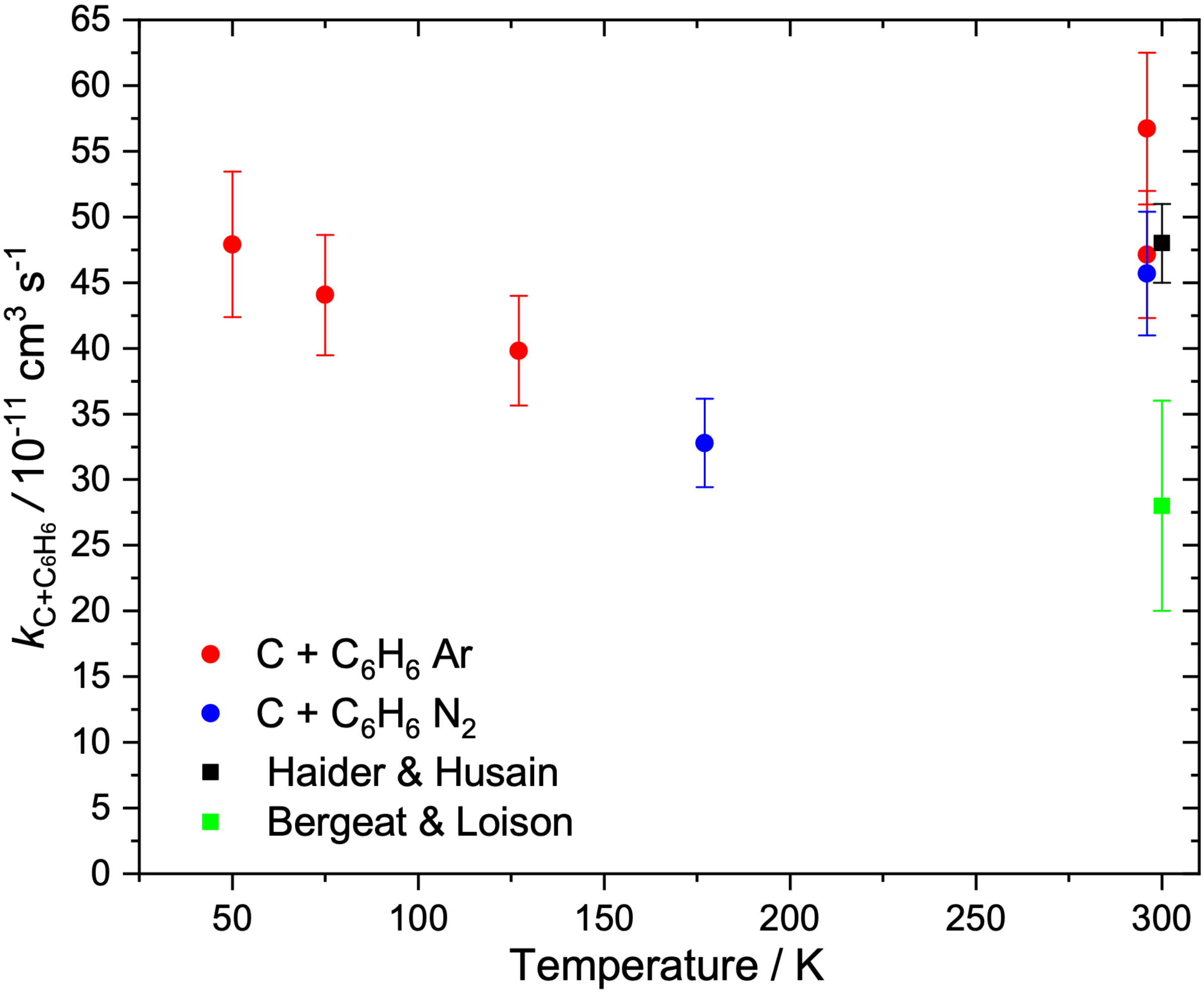

**Figure 3** Temperature dependence of the measured second-order rate constants for the C + $C_6H_6$ reaction. (Red circles) this work, argon carrier gas; (blue circles) this work, nitrogen carrier gas; (black square) Haider and Husain;[20] (green square) Bergeat and Loison.[25] Errors on the values derived in the present work represent the combined statistical and systematic uncertainties. Statistical uncertainties were obtained from weighted linear-least squares fits such as those shown in Figure 3. Systematic uncertainties were estimated to be 10 % of the nominal value of the second-order rate constant.

At 296 K, several experiments were performed to check for a possible pressure dependence and any eventual dependence on the nature of the carrier gas. The rate constant obtained with Ar as the carrier gas at a flow density of 1.63 $\times$ $10^{17}$ $cm^{-3}$ is smaller (4.71 $\times$ $10^{-10}$ $cm^3$ $s^{-1}$) than the rate constant obtained at a flow density of 6.7 $\times$ $10^{16}$ $cm^{-3}$ (5.67 $\times$ $10^{-10}$ $cm^3$ $s^{-1}$) although the combined error bars of the individual measurements overlap. Individual decays were also recorded at even higher flow densities (3.26 and 6.5 $\times$ $10^{17}$ $cm^{-3}$) for a fixed [$C_6H_6$] = 1.34 $\times$ $10^{14}$ $cm^{-3}$, where the derived $k_{1st}$ values were consistent with those obtained in experiments at a lower flow density (1.63 $\times$ $10^{17}$ $cm^{-3}$) with similar $C_6H_6$ concentrations.

The rate constant measurements with $N_2$ as the carrier gas at flow densities of 1.63 $\times$ $10^{17}$ $cm^{-3}$ and 0.66 $\times$ $10^{17}$ $cm^{-3}$ resulted in second-order plots with very similar slopes and intercept values, so these data were combined to give an overall rate constant of 4.57 $\times$ $10^{-10}$ $cm^3$ $s^{-1}$ which is just outside of the estimated error bars of the lower flow density Ar experiment. The values determined in the present study, ranging from (4.6-5.7) $\times$ $10^{-10}$ $cm^3$ $s^{-1}$ are in good agreement with the previous work of Haider and Husain who measured a rate constant of (4.8 $\pm$ 0.3) × $10^{-10}$ $cm^3$ $s^{-1}$ at room temperature.[20] Pressures in the range 1-50 Torr were used here, with He as the carrier gas. The rate constant value derived in the previous work of Bergeat and Loison[25] of (2.8 $\pm$ 0.8) × $10^{-10}$ $cm^3$ $s^{-1}$, who used a total pressure of 2 Torr with He as the carrier gas, is somewhat lower than those obtained in the present study and in the work of Haider and Husain.[20]

At lower temperature, it can be seen that the rate constant derived from the measurement at 177 K with $N_2$ as the carrier gas is somewhat lower than those obtained at other temperatures with values in the range (4.0-5.7) × $10^{-10}$ $cm^3$ $s^{-1}$. It is difficult to establish a firm explanation for this difference due to the challenges associated with varying the pressure in Laval nozzle experiments, although one possible reason could be the possible formation of van der Waals complexes between $C_6H_6$ and $N_2$ at low temperature which might lower the gas-phase free benzene concentration resulting in an underestimation of the rate constant. In this respect, previous work[33] has shown that $C_6H_6$ and $N_2$ forms a complex that is at least twice as strongly bound as the one between $C_6H_6$ and Ar, with $N_2$ lying parallel to the benzene molecule.[34] Indeed, earlier studies of the low temperature kinetics of the reactions of $C_6H_6$ with various radical species seem to indicate that those experiments performed with $N_2$ as the carrier gas generally furnish rate constants that are slightly lower than the equivalent ones where noble gases were used as the carrier gas (see for example Table 1 of Cooke et al. [8]). As a result, this rate constant was not used to derive the temperature dependence of the C + $C_6H_6$ reaction. As the other measured rate constants are mostly within the combined measurement uncertainties, we recommend the use of a temperature independent value for the rate constant, equal to $k_{C+C_6H_6} = (4.69 \pm 0.52) \times 10^{-10}$ cm$^3$ s$^{-1}$, derived from an average of the remaining individual second-order rate constants determined here. The error bars on this value represent the derived statistical uncertainty with a 10 % systematic uncertainty added.

### 3.2 H-atom product yields

The measurement of H-atom yields allows us to provide some information regarding the product channels of the C + $C_6H_6$ reaction. In this work, the formation of atomic hydrogen was followed as a function of time by VUV LIF at 121.567 nm. The measured intensities were calibrated by comparison with the H-atom yield of a reference reaction, namely C + $C_2H_4$, with a known H-atom yield of 0.92 ± 0.04 at 300 K.[25] The H-atom yield of the C + $C_2H_4$ reaction was assumed not to vary as a function of temperature as explained previously. [32]

Earlier measurements of the absorption cross-sections of $C_6H_6$ at UV wavelengths[35] have shown that benzene absorbs only weakly around 266 nm, with a value of approximately $2 \times 10^{-20}$ cm$^{-2}$. Despite this, significant quantities of atomic hydrogen were generated in preliminary experiments by $C_6H_6$ photolysis at this wavelength in the absence of atomic carbon. This is thought to arise from multiphoton absorption effects as a single photon at 266 nm is not energetic enough to break the C-H bond of $C_6H_6$.[36] Consequently, two different experiments were performed, allowing us to discriminate between the photolysis signal and the reactive signal, that is, the H-atom signal produced by the C + $C_6H_6$ reaction. The results of a typical set of experiments performed at 296 K with a total flow density of $1.6 \times 10^{17}$ cm$^{-3}$ with $N_2$ as the carrier gas are shown in Figure 4. Experiments were performed at lower ($6.8 \times 10^{16}$ cm$^{-3}$) and higher ($3.2 \times 10^{17}$ cm$^{-3}$) flow densities at 296 K to check for the possible influence of pressure on the H-atom yields and also at 177 K to examine the temperature dependence of the H-atom yields. Each set of data points shown in Figure 4 represent the results of 5 coadded temporal scans.

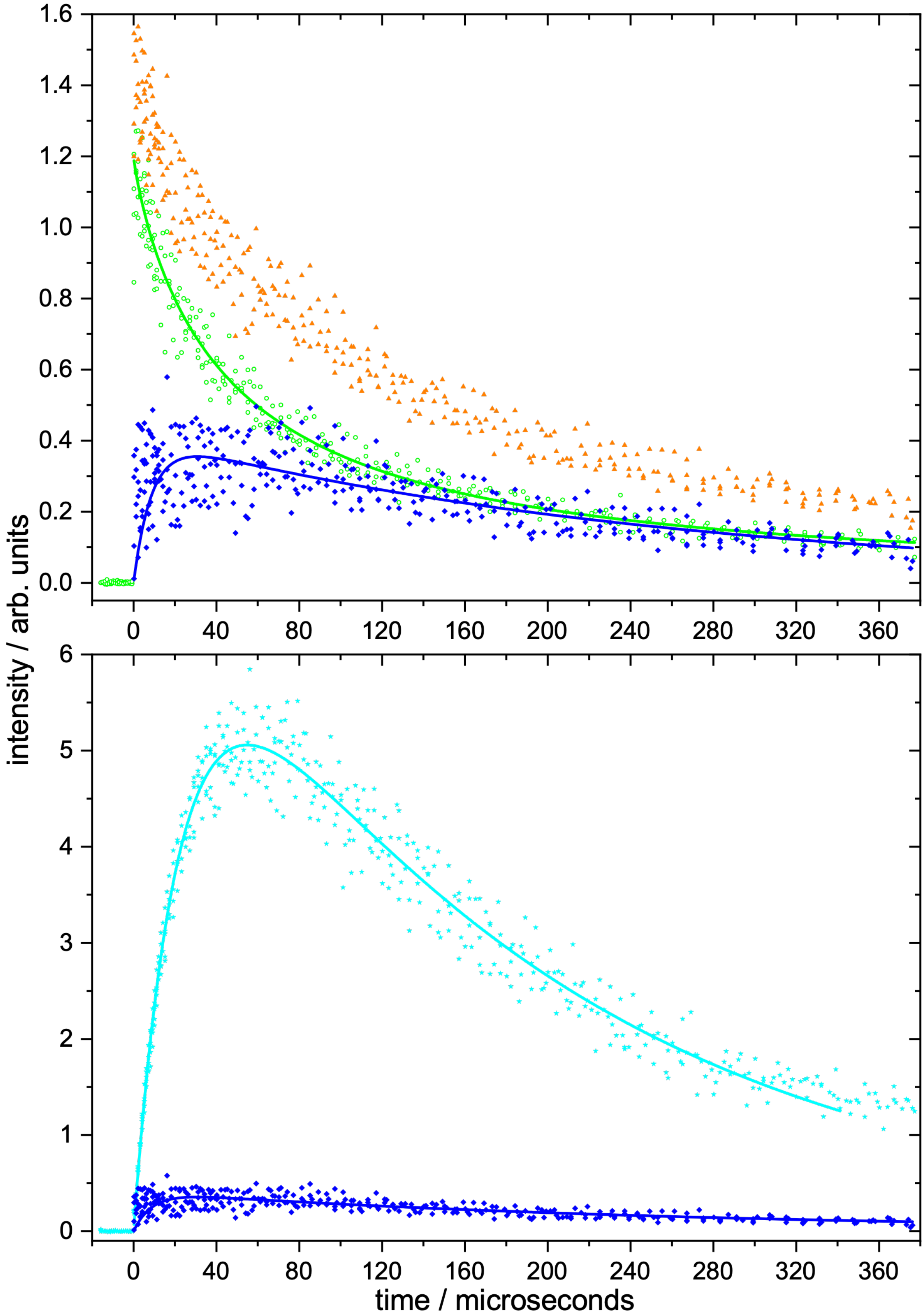

**Figure 4** Atomic hydrogen fluorescence profiles recorded at 296 K. **Upper panel** (Green circles) H-atom signal in the absence of atomic carbon with $C_6H_6$ present in the reactor with $[C_6H_6]$ =

$1.92 \times 10^{13}$ cm$^{-3}$; (green line) arbitrary fit to the green data set; (orange triangles) H-atom signal with both atomic carbon and $C_6H_6$ present in the reactor with [$C_6H_6$] = $1.9 \times 10^{13}$ cm$^{-3}$; (blue diamonds) subtraction of the fit to the green data set from the orange data set, corresponding to the reactive signal. **Lower panel** (cyan stars) H-atom signal from the C + $C_2H_4$ reference reaction with [$C_2H_4$] = $4.1 \times 10^{13}$ cm$^{-3}$; (blue diamonds) C + $C_6H_6$ reactive signal. The solid blue and cyan lines represent biexponential fits to the corresponding data sets using expression E3.

In the upper panel, the orange data points show the H-atom signal intensity as a function of time recorded with both $CBr_4$ (and therefore C-atoms following photolysis) and $C_6H_6$ present in the reactor. Under these conditions, this temporal trace represents the sum of both the photolysis and reactive contributions to the H-atom signal. The green data points show the results of a second experiment where the $CBr_4$ flow has been turned off, thereby eliminating the reactive contribution to the H-atom signal due to the absence of atomic carbon. A fit to the green data points by an arbitrary function allows us to remove the photolysis signal from the photolysis + reactive one, with the blue data points showing the remaining reactive signal following subtraction. In the lower panel, we show the H-atom signal obtained from the reference C + $C_2H_4$ reaction as the cyan data points alongside the blue reactive signal datapoints obtained for the C + $C_6H_6$ reaction as described above.

Both the blue and cyan data points are well described by the expected biexponential expression for H-atom formation

$$I_H = A\{\exp(-k_{L(H)}t) - \exp(-k_{1st}t)\} \quad \text{E3}$$

where $I_H$ is the time dependent H-atom intensity, $k_{L(H)}$ is the rate constant for H-atom loss, $k_{1st}$ is the rate constant for H-atom formation and A is the H-atom signal amplitude. H-atom losses are considered to arise mostly from diffusion out of the observation region. The H-atom yield for the C + $C_6H_6$ reaction is derived by comparing the peak intensity values of the biexponential fits to these two data sets. The final H-atom yield is then derived by correcting for the fact that the absolute H-atom yield of the reference reaction is 0.92 rather than 1. No correction was required for absorption of the VUV excitation source by $C_6H_6$ or $C_2H_4$ as the low concentrations of these reagents was estimated to lead to negligible absorption losses. The final H-atom yield derived from the experiments shown in Figure 4 is 6.5 %, with similar values of 8.1 % and 5.8 % obtained at higher and lower pressures respectively. For the measurements performed at the lowest pressure ($[N_2] = 6.8 \times 10^{16}$ cm$^{-3}$) the subtraction yielded a reactive curve with a non-zero value following photolysis. Under these conditions, $C(^1D)$ atoms produced by $CBr_4$ photolysis are less efficiently quenched by $N_2$[37] with an estimated half-life time of approximately 2 μs. Consequently, it is likely that these additional H-atoms arise from the $C(^1D) + C_6H_6$ reaction. For these curves, rather than using the peak value of the H-atom intensity (in this case at $t$ = 0 μs) we used the H-atom intensity value at 50 μs to establish the H-atom yield, although this is still likely to contain a contribution from H-atom production by the $C(^1D) + C_6H_6$ reaction, so that the real value of the H-atom yield for this measurement may be even lower. While these results suggest that the H-atom yield increases with pressure, supporting the idea of a possible pressure dependence, it is almost certainly inappropriate to overinterpret the data in this instance. Indeed, the H-atom yields themselves are derived from the subtraction of two much larger signals, and therefore subject to large uncertainties. Consequently, we prefer to restrict our discussion of the H-atom yields derived at 296 K to the overarching result which is that the H-atom yields are clearly very low at this temperature, in reasonably good agreement with the earlier work of Bergeat and Loison who determined a H-atom yield of $0.16 \pm 0.06$ at 300 K.[25] At 177 K, where it was not possible to vary the pressure, we obtain a similarly low H-atom yield of 6.3 %, indicating the probable weak temperature dependence of the H-atom yield. These results are in contrast to the results of crossed-molecular beam studies of the $C + C_6H_6/C_6D_6$ reactions reported by the Kaiser group[22, 23] who suggested that the reaction should lead to the formation of fulvenallenyl, $C_7H_5$ + H ($C_7D_5$ + D) as products. It should be noted here that these authors also detected the adduct $C_7D_6$ in experiments on the $C + C_6D_6$ reaction. As these measurements are conducted under single collision conditions, the direct detection of $C_7D_6$ indicates that this species is particularly stable with a long lifetime, despite its high internal energy of approximately 300 kJ/mol. Hahndorf et al.[23] also stated that intersystem crossing (ISC) to the singlet state was likely to be only a minor pathway in these reactions as the products $C_7H_4$ (+ $H_2$) should have been observed in experiments performed at higher collision energies if ISC was important at the collision energies used in their experiments. However, as noted by Hahndorf et al.[23] the lifetime of the triplet $C_7H_6$ adducts are estimated to be between $10^{-12}$ s and $2.5\times10^{-4}$ s for their experimental conditions but this drastically increases at lower energies approaching 1 s at very low energy for one of them. Therefore, the experimental results obtained in the crossed molecular beam experiments at high energy are not necessarily valid at room temperature or below, let alone for the very low energies corresponding to the conditions of the interstellar medium, where ISC could become significant. Consequently, it is difficult to determine if ISC is responsible for the discrepancy between the crossed molecular beam experiments and the results obtained here. Additionally, it is not clear whether other possible product channels such as those leading to c-$C_5H_4$ + $C_2H_2$ were examined experimentally during this course of these measurements, so it remains a possibility that these products were also formed during these earlier studies and not detected. Despite this, the possible formation of c-$C_5H_4$ + $C_2H_2$ as products was discussed by Bettinger et al.[38] while Hahndorf et al.[23] calculated a TS of 16 kJ/mol relative to the reagents for the dissociation step from intermediate cycloheptatrienylidene to c-$C_5H_4$ + $C_2H_2$. A more recent study by da Silva[26] has calculated this TS to be significantly below the reagent energy level (-43 kJ/mol) using the composite G3X-K quantum chemical method.

Comparing our work with that of Kaiser's group is difficult not only due to the difference in energy within the system but also because of the difference in collision conditions. Indeed, under our current experimental conditions of high densities ($10^{17}$ $cm^{-3}$) and multiple collisions it is possible that at least some of the $C_7H_6$ adducts formed by the C + $C_6H_6$ reaction are stabilized, leading to a corresponding decrease in the yield of all bimolecular product channels. At the present time, there is little or no experimental evidence to rule out the formation of c-$C_5H_4$ + $C_2H_2$ as products, although this is still expected to be a minor pathway considering the exothermicities of the c-$C_5H_4$ + $C_2H_2$ and $C_7H_5$ + H exit channels and the predicted heights of the TSs[26] leading to these products.

## 4 Astrochemical model

The reaction between C and $C_6H_6$ is not currently included in astrochemical databases such as KIDA[12] or UMIST.[39] Despite this, it is likely to play an important role in astrochemical models due to the high estimated abundance of atomic carbon. In the recent study by Loison et al.,[14] the C + $C_6H_6$ reaction was included with an estimated rate constant and was found to represent the main consumption pathway for $C_6H_6$. In their study, Loison et al.[14] estimated a temperature independent rate constant of $4.0 \times 10^{-10}$ $cm^3$ $s^{-1}$ for this process, in good agreement with the rate constant measured in this work of $k_{\mathrm{C+C_6H_6}} = (4.69 \pm 0.52) \times 10^{-10}$ $cm^3\ s^{-1}$. They assumed a 100% yield of c-$C_5H_4CCH$ + H as products based on the theoretical study by da Silva.[26] Given the results of our study and the significant uncertainties regarding the products and branching ratios, we examined the effects of several scenarios, based on the different experimental measurements and theoretical calculations, on astrochemical simulations of dense molecular clouds using the Nautilus code.[11] The Nautilus code[11] is a 3-phase gas, dust grain ice surface and dust grain ice mantle time dependent chemical model employing kida.uva.2024[40] as the basic reaction network updated recently for a better description of complex organic molecule (COM) chemistry on interstellar dust grains and in the gas-phase[41-44] and including the recent updates of Loison et al.[14] There are 800 individual species included in the network that are involved in approximately 9000 separate reactions. The C/O gas phase elemental ratio is equal to 1 in this work. This ratio, which is higher than the cosmic abundance ratio (0.6), allows for a much better agreement between the model and the entire set of observations for dense molecular clouds such as TMC-1. Otherwise, there is not enough carbon to consume sufficient oxygen atoms (leading to CO), and the remaining oxygen atoms limit the abundances of radicals, thereby inhibiting the chemistry. This is also the choice made by Hincelin et al.[45] and already used in their study of $O_2$ as well as in recent studies of aromatic species and in their complex carbon chemistry.[46-48] Physically, this oxygen depletion can be explained by the formation of water ice during the history of the dense cloud—water ice that does not desorb in the warmer less dense phases of evolution, unlike CO[49, 50] which dissociates into C and O in this less dense phase. The initial abundances used are given in Table 2. The total density (nH + 2n$H_2$) is equal to $5.0 \times 10^4$ $cm^{-3}$, the temperature is equal to 10 K, the ionization by cosmic ray is equal to $1.3 \times 10^{-17}$ $s^{-1}$, the visual extinction is equal to 10.

**Table 2 Elemental abundances**

| Element | Abundance[(a)] |
|---|---|
| $H_2$ | 0.5 |
| He | 0.09 |
| $C^+$ | $1.4 \times 10^{-4}$ |
| N | $3.0 \times 10^{-5}$ |
| O | $1.4 \times 10^{-4}$ |

| $s\text{-}H_2O^{(b)}$ | $1.0 \times 10^{-4}$ |
|---|---|
| $S^+$ | $4.0 \times 10^{-6}$ |
| $Fe^+$ | $2.0 \times 10^{-8}$ |
| $Cl^+$ | $1.0 \times 10^{-7}$ |
| F | $6.7 \times 10^{-9}$ |

[(a)]Relative to total hydrogen nuclei

[(b)] s-$H_2O$ means $H_2O$ on grains (surface and bulk)

The grain surface and the mantle are both chemically active for these simulations, while accretion and desorption are only allowed between the surface and the gas-phase. The dust-to-gas ratio (in terms of mass) is 0.01. A sticking probability of 1 is assumed for all neutral species while desorption occurs by both thermal and non-thermal processes (cosmic rays, chemical desorption) including sputtering of ices by cosmic-ray collisions.[51] A full description of the surface reaction formalism and simulations is available in Ruaud et al.[11]

**Scenario 1.**

In this case, we consider that the branching ratios obtained in the Laval nozzle experiments are valid at low pressure and that the reaction remains on the triplet surface. Therefore, we assume that the major pathway here leads to c-$C_5H_4$ + $C_2H_2$ products, with only a minor channel leading to atomic hydrogen and $C_7H_5$. While other $C_5H_4$ isomers exist that are more stable than c-$C_5H_4$ (such as $CH_3C_4H$ and $H_2C_3HC_2H$), they are all characterized by a singlet ground state. c-$C_5H_4$ has a triplet ground state and is the only $C_5H_4$ isomer which can be produced on the triplet C + $C_6H_6$ surface.

$$C + C_6H_6 \rightarrow C_7H_5 + H \ (15\%)$$

$$\rightarrow c\text{-}C_5H_4 + C_2H_2 \ (85\%)$$

**Scenario 2.**

In this case, we consider that, under interstellar medium conditions—corresponding to very low collision energies—the lifetime of the $C_7H_6$ adduct is sufficiently long to favor stabilization through photon emission (after ISC) by comparing with similar neutral systems such as $CH_3$ + $C_6H_5$[52] and many ionic ones such as $C_4H_3^+$ + $C_2H_2$, $C_5H_5^+$ + $C_2H_2$.[53] In this case, there is very little production of c-$C_5H_4$ + $C_2H_2$. Indeed, according to the calculations of da Silva,[26] it seems difficult to imagine a high production of c-$C_5H_4$ + $C_2H_2$, which could explain the low experimental production of atomic hydrogen in Laval nozzle and flow reactor experiments, without also favoring a high production of H-atoms. In the Laval nozzle and flow reactor experiments, due to the high flow densities ($10^{16}$-$10^{17}$ $cm^{-3}$) three body collisions would stabilize the $C_7H_6$ adduct, playing a role similar to the one played by radiative stabilization at very low pressure.

$C + C_6H_6 \rightarrow C_7H_5 + H$ (15%)

$\rightarrow C_7H_6(c\text{-}C_5H_4CCH_2) + h\nu$ (85%)

**Scenario 3:**

In this case we consider that the H-atom yields obtained in the Laval nozzle and flow reactor experiments are influenced by three-body collisions, while radiative stabilization is inefficient. Then under the conditions of very low pressure relevant to the interstellar medium, the pathway leading to H + $C_7H_5$ products accounts for 100% of the reaction, in a similar manner to the results of high-energy crossed molecular beam experiments.

$C + C_6H_6 \rightarrow C_7H_5 + H$ (100%)

**Scenario 4.**

We exclude the C + $C_6H_6$ reaction from the network.

In Figure 5, we compare the models of the various scenarios with the observed abundances (when they exist) for c-$C_5H_6$, $C_6H_4$, $C_6H_6$, $C_7H_6$ (sum of all isomers), $C_6H_5CN$ and $C_6H_5C_2H$ produced in the four different scenarios, and compare these to the observed abundances where they exist.

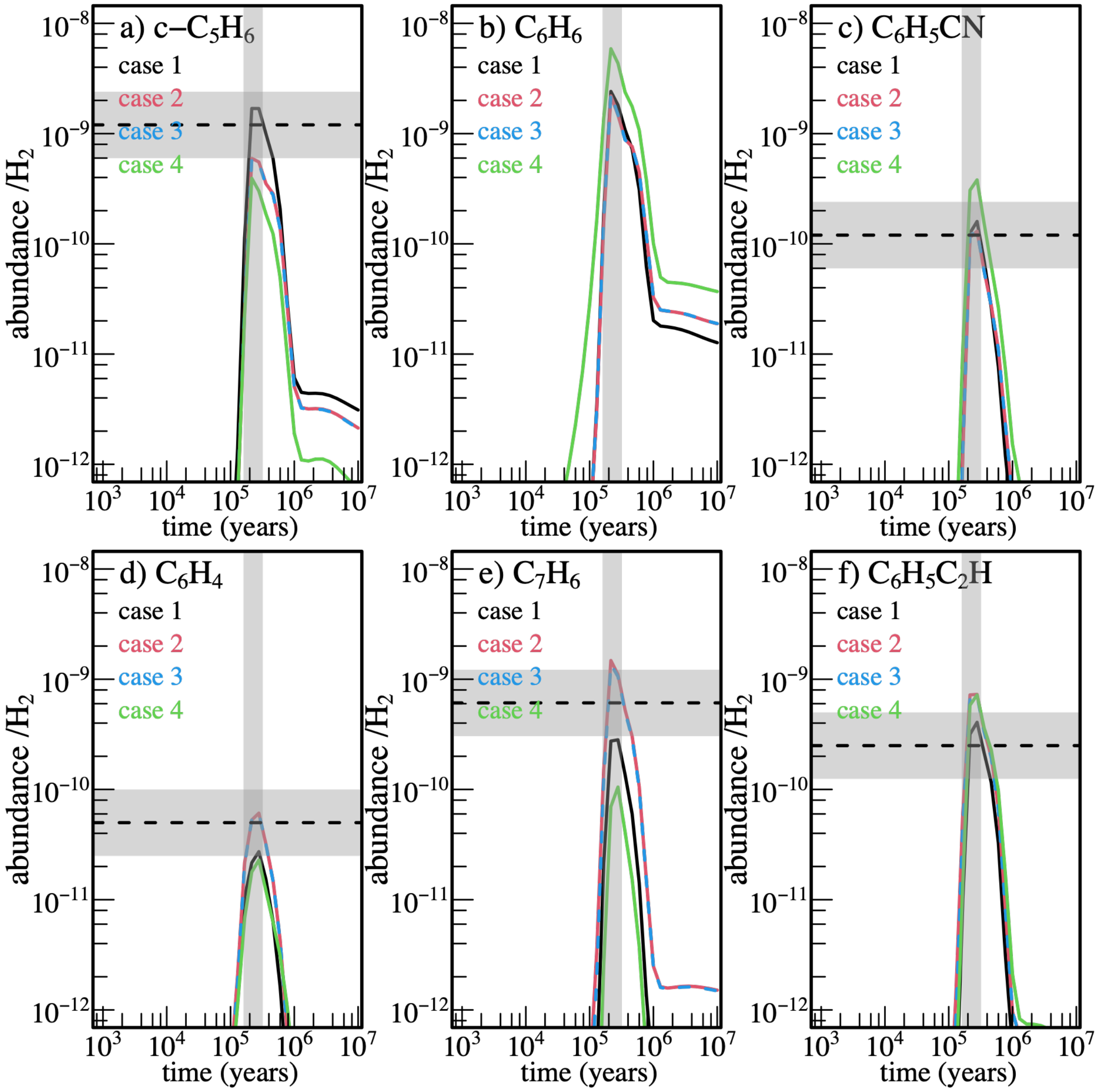

**Figure 5** Gas-grain astrochemical model results for c-$C_5H_6$, $C_6H_4$, $C_6H_6$, $C_6H_5CN$, $C_6H_5C_2H$ formation in a dark cloud as a function of time. (Solid black line) scenario 1; (Solid red line) scenario 2; (dashed blue line) scenario 3; (Solid green line) scenario 4. The horizontal rectangle represents the observed abundances in TMC-1 (given by the horizontal dashed lines) with an arbitrary error associated (±2): c-$C_5H_6$,[54] $C_6H_4$,[55] $C_7H_6$ (sum of c-$C_5H_4CCH_2$[56] + 1-c-$C_5H_5CCH$[57] + 2-c-$C_5H_5CCH$[57]), $C_6H_5CN$,[7] $C_6H_5C_2H$.[58]

It can be seen that for benzene and its direct derivatives, $C_6H_5CN$ and $C_6H_5C_2H$, there are only minor differences between the first three scenarios and a good agreement with observations for $C_6H_5CN$[7] and $C_6H_5C_2H$,[58] even if scenarios 2 and 3 slightly overestimate the $C_6H_5C_2H$ abundance. This is logical because, since the consumption rate of $C_6H_6$ is unchanged, its

abundance—and consequently the abundances of $C_6H_5CN$ and $C_6H_5C_2H$ produced by the reactions CN + $C_6H_6$ and $C_2H$ + $C_6H_6$—remains largely unaffected. The abundance of $C_6H_5C_2H$ changes slightly in scenario 1, as $C_6H_5C_2H$ is produced not only by the $C_2H$ + $C_6H_6$ reaction, but also by the C + $C_7H_6$ reaction. In scenario 1, $C_7H_6$ is produced less efficiently, resulting in a lower abundance of $C_6H_5C_2H$.

Removing the C + $C_6H_6$ reaction induces slightly more change. In this case, the abundance of $C_6H_6$ increases significantly, as do its derived products—particularly $C_6H_5CN$, which is produced solely by the CN + $C_6H_6$ reaction. For $C_6H_5C_2H$, its abundance decreases because, although the abundance of $C_6H_6$ increases, making the $C_2H$ + $C_6H_6$ reaction more efficient, less $C_7H_6$ is produced. This, in turn, is unfavorable for the pathway involving $C_6H_5C_2H$ production by the C + $C_7H_6$ reaction. In dense clouds, closed-shell species such as $C_6H_6$ are expected to react efficiently only with highly reactive species such as C, CH, $C_2H$, CN and with ions such as $C^+$, $H_3^+$ and $HCO^+$. Considering the relative abundances, the most important reactions are with C and with $C^+$, $H_3^+$ and $HCO^+$. $C^+$ reacts mostly with $C_6H_6$ through charge exchange[59] and $H_3^+$ and $HCO^+$ through $H^+$ transfer. Then these ionic reactions are not expected to constitute a significant destruction pathway as $C_6H_6^+$ and $C_6H_7^+$ are expected to recombine efficiently with electrons leading mostly to $C_6H_5$ and $C_6H_6$, thereby regenerating benzene.[60] Consequently, the C + $C_6H_6$ reaction is the most efficient one for benzene loss, explaining why its removal induces notable changes.

The abundance of $C_6H_4$ is in better agreement with scenarios 2 and 3 because we consider that the C + $C_7H_6$ reaction leads to some $C_6H_4$ + $C_2H_2$ which becomes an important source of $C_6H_4$.

For c-$C_5H_6$, it can be seen that scenario 1, which involves significant production of c-$C_5H_4$ through the C + $C_6H_6$ reaction, yields much more c-$C_5H_6$ with a better agreement with observations. This is because the c-$C_5H_4$ produced will undergo barrierless reactions with atoms, primarily with atomic hydrogen, producing c-$C_5H_5$ through radiative association, and is therefore an alternative production pathway for c-$C_5H_6$, which is currently underestimated in astrochemical models. [48, 61] It is worth noting, however, that the c-$C_5H_6$ abundance simulated by our revised model is close to the observed one. This is because, compared to our previous study,[61] we have explored additional sources of c-$C_5H_6$. In our previous study, c-$C_5H_6$ was produced by the following processes:

-$C_2H_4^+$ + $CH_3CCH$ and $C_3H_7^+$ + $C_2H_2$ reactions followed by Dissociative Electronic Recombination (DR) of c-$C_5H_7^+$, representing approximately 10% of the observed abundance

-CH + $C_4H_6$, representing approximately 1% of the observed abundance

-O + $C_6H_5$ + H and C + $C_4H_6$ +H + H, representing approximately 10% of the observed abundance

We have added the following pathways, all involving c-$C_5H_5$ production:

- CH + $C_6H_6$ $\rightarrow$ c-$C_5H_5$ + $C_2H_2$ using the rate constant measurement from Hamon et al.[17] and estimating the branching ratio from the theoretical study of He et al.,[62] representing approximately 10% of the observed abundance

- DR of $C_7H_7^+$ (tropylium and benzylium ions) which produces in the first step neutral $C_7H_7$ and thus we end up on the same potential energy surface as the CH + $C_6H_6$ reaction, representing approximately 40% of the observed abundance of c-$C_5H_5$ with a branching ratio $C_7H_7^+$ + $e^-$ $\rightarrow$ c-$C_5H_5$ + $C_2H_2$ equal to 30%.

By using an 85% branching ratio for the C + $C_6H_6$ $\rightarrow$ c-$C_5H_4$ + $C_2H_2$ pathway, we obtain a slightly higher simulated c-$C_5H_6$ abundance than the observed one. Given the various uncertainties, a slightly smaller branching ratio is plausible and would allow for excellent agreement between the model and observations for this species.

The most significant changes across the different scenarios are observed for $C_7H_6$. There are several isomers of $C_7H_6$, with c-$C_5H_4CCH_2$ being the most stable, three of them being observed in dense molecular clouds (c-$C_5H_4CCH_2$,[56] 1-c-$C_5H_5CCH$[57] and 2-c-$C_5H_5CCH$[57]). In our network, we merge all isomers, particularly c-$C_5H_4CCH_2$ (fulvenallene, potentially produced through the C + $C_6H_6$ reaction[26]) and 1-c-$C_5H_5CCH$ and 2-c-$C_5H_5CCH$ (potentially produced through the c-$C_5H_6$ + $C_2H$ reaction). It is worth noting that the 7-carbon ring cyclic $C_7H_6$ isomer, that could also be produced by the C + $C_6H_6$ reaction according to theoretical calculations,[26, 38] is a carbene and likely exhibits much higher reactivity than the closed-shell molecules c-$C_5H_4CCH_2$ and c-$C_5H_5CCH$ especially with atomic hydrogen, causing isomerization into the most stable c-$C_5H_4CCH_2$ isomer.[62] It can be noted that scenarios 2 and 3 give very similar results because the $C_7H_5$ produced in scenario 3 (potentially various isomers) is a radical and will primarily react with atoms—mainly H atoms in dense molecular clouds. Here the H + $C_7H_5$ reaction occurs on the singlet potential energy surface, in contrast to the C + $C_6H_6$ reaction where reaction occurs over the triplet surface. In this case, either there will be isomerization toward the most stable $C_7H_5$ isomer or stabilization through photon emission. Ultimately, $C_7H_6$ will stabilize through photon emission since there will no longer be any exothermic pathways leading to bimolecular products. The most likely $C_7H_6$ isomer formed should be the most stable one, c-$C_5H_4CCH_2$ (fulvenallene).[26] So the $C_7H_5$ produced by the C + $C_6H_6$ reaction will mostly lead to $C_7H_6$ formation. Depending on the scenario considered, $C_7H_6$ can reach relatively high abundances. In Figure 5, we compared the calculated abundance of $C_7H_6$ with the sum of the abundances of the three observed isomers. Scenarios 2 and 3, which favor high production of $C_7H_5$ and $C_7H_6$, slightly overestimate the abundance of $C_7H_6$, while Scenario 1, which favors the production of c- $C_5H_4$ + $C_2H_2$, slightly underestimates the abundance of $C_7H_6$. Scenario 4, which neglects the C + $C_6H_6$ reaction, strongly underestimates the abundance of $C_7H_6$, highlighting the critical role of this reaction (in this case, $C_7H_6$ is produced by $C_2H$ + c- $C_5H_6$). It can also be noted that the best agreement for $C_7H_6$ isomers and c-$C_5H_6$ is when the C + $C_6H_6$ $\rightarrow$ c-$C_5H_4$ + $C_2H_2$ and C + $C_6H_6$ $\rightarrow$ $C_7H_5$ + H/ $C_7H_6$ + h$\nu$ pathways are non negligeable.

A key aspect of aromatic chemistry in the ISM involves the reactions of atomic hydrogen, particularly radiative associations. At typical ages of dense clouds, atomic hydrogen is significantly more abundant than other species, except for $H_2$, because even though it depletes onto grains like all other species, it is continuously produced by the reaction $H_2^+$ + $H_2$ (with $H_2^+$ being generated by the interaction of $H_2$ with cosmic rays). Radiative associations are important because the reactions involved (H + c-$C_5H_5$, H + $C_6H_5$) lack exothermic bimolecular exit channels, and the sizes of the systems allow for a sufficiently long lifetime before back-dissociation occurs (dissociation to reform the reactants), allowing stabilization to occur by photon emission. It should be noted also that radiative association reactions between neutrals are not well characterized, particularly at the experimental level. To estimate their values, we rely on theoretical studies.[52, 63]

**5 Conclusions**

This work presents a combined experimental and astrochemical modeling study of the C + $C_6H_6$ reaction. Experiments were performed using a continuous supersonic flow reactor employing pulsed laser photolysis for C-atom generation and laser induced fluorescence to follow C-atom decay in the presence of coreagent $C_6H_6$, allowing the reaction kinetics to be studied over the 50-296 K temperature range. The rate constants were found to be large and independent of temperature over this range, highlighting the barrierless nature of the underlying triplet potential energy surface, in good agreement with previous theoretical work. Experiments were also performed to determine the H-atom yield of the reaction at 296 K and 177 K, by comparison with the H-atom yield of the C + $C_2H_4$ reaction. These measurements showed that H-atom yields are low (< 10 %) at both temperatures, suggesting that the primary products could be c-$C_5H_4$ + $C_2H_2$ under the present experimental conditions based on earlier calculations or alternatively $C_7H_6$ by collisional stabilization of the adduct. The effects of this reaction on the abundances of several related carbon bearing species were tested by including it in a gas-grain dense cloud model, recently updated for a better description of COM chemistry. Several different scenarios were considered during these simulations to account for the contrasting results of the various experimental studies of the product formation channels. Given the many uncertainties and the differences between the experiments conducted and the interstellar medium (our experiments were carried out at low temperatures but at much higher pressures than those of the interstellar medium, and crossed-beam experiments were performed at low pressures but at much higher energies than those characterizing the interstellar medium), it is difficult to favor one of the proposed scenarios. Nevertheless, the model results show that the C + $C_6H_6$ reaction is the major loss process for interstellar benzene, while the best agreement with observations of related interstellar species (c-$C_5H_6$, $C_6H_5CN$ and $C_6H_5C_2H$) is obtained when H-atom producing channels are considered to be only a minor pathway of the C + $C_6H_6$ reaction. This work also suggests that the C + $C_6H_6$ reaction could be a non-negligible production pathway to $C_7H_6$ (c-$C_5H_4CCH_2$, fulvenallene) which can reach relatively high abundances.

**Acknowledgements**

K. M. H. acknowledges support by the thematic action “Physique et Chimie du Milieu Interstellaire” (PCMI) of the INSU National Programme “Astro”, with contributions from CNRS Physique & CNRS Chimie, CEA, and CNES. K. M. H. also acknowledges support from the ‘‘Programme National de Planétologie’’ (PNP) of the CNRS/INSU.

**References**

(1) Tielens, A. G. G. M. Interstellar Polycyclic Aromatic Hydrocarbon Molecules. *Annu. Rev. Astron.* **2008**, *46*, 289-337.

(2) Burkhardt, A. M.; Loomis, R. A.; Shingledecker, C. N.; Lee, K. L. K.; Remijan, A. J.; McCarthy, M. C.; McGuire, B. A. Ubiquitous Aromatic Carbon Chemistry at the Earliest Stages of Star Formation. *Nat. Astron.* **2021**, *5*, 181-187. DOI: 10.1038/s41550-020-01253-4.

(3) Cernicharo, J.; Heras, A. M.; Tielens, A. G. G. M.; Pardo, J. R.; Herpin, F.; Guelin, M.; Waters, L. B. F. M. Infrared Space Observatory's Discovery of $C_4H_2$, $C_6H_2$, and Benzene in CRL 618. *The Astrophys. J. Lett.* **2001**, *546*, L123-L126.

(4) Coustenis, A.; Salama, A.; Schulz, B.; Ott, S.; Lellouch, E.; Encrenaz, T. h.; Gautier, D.; Feuchtgruber, H. Titan's Atmosphere from ISO Mid-Infrared Spectroscopy. *Icarus* **2003**, *161*, 383-403. DOI: https://doi.org/10.1016/S0019-1035(02)00028-3.

(5) Vuitton, V.; Yelle, R. V.; Cui, J. Formation and Distribution of Benzene on Titan. *J. Geophys. Res.* **2008**, *113*, E05007. DOI: 10.1029/2007je002997.

(6) Arabhavi, A. M.; Kamp, I.; Henning, Th.; van Dishoeck, E. F.; Christiaens, V.; Gasman, D.; Perrin, A.; Güdel, M.; Tabone, B.; Kanwar, J.; et al. Abundant Hydrocarbons in the Disk around a Very-Low-Mass Star. *Science*, **2024**, *384*, 1086-1090.

(7) McGuire, B. A.; Burkhardt, A. M.; Kalenskii, S.; Shingledecker, C. N.; Remijan, A. J.; Herbst, E.; McCarthy, M. C. Detection of the Aromatic Molecule Benzonitrile (C-$C_6H_5CN$) in the Interstellar Medium. *Science* **2018**, *359*, 202-205. DOI: 10.1126/science.aao4890.

(8) Cooke, I. R.; Gupta, D.; Messinger, J. P.; Sims, I. R. Benzonitrile as a Proxy for Benzene in the Cold ISM: Low-Temperature Rate Coefficients for CN + $C_6H_6$. *Astrophys. J. Lett.* **2020**, *891*, L41. DOI: 10.3847/2041-8213/ab7a9c.

(9) Trevitt, A. J.; Goulay, F.; Taatjes, C. A.; Osborn, D. L.; Leone, S. R. Reactions of the CN Radical with Benzene and Toluene: Product Detection and Low-Temperature Kinetics. *J. Phys. Chem. A* **2010**, *114*, 1749-1755. DOI: 10.1021/jp909633a.

(10) Xue, C.; Byrne, A. N.; Morgan, L.; Wenzel, G.; Changala, P. B.; Fried, Z. T. P.; Loomis, R. A.; Remijan, A.; Bergin, E. A.; Cooke, I. R.; et al. The Molecular Inventory of TMC-1 with GOTHAM Observations. *Astrophys. J., Suppl. Ser.* **2025**, *281*, 9.

(11) Ruaud, M.; Wakelam, V.; Hersant, F. Gas and Grain Chemical Composition in Cold Cores as Predicted by the Nautilus Three-Phase Model. *Mon. Not. R. Astron. Soc.* **2016**, *459*, 3756-3767. DOI: 10.1093/mnras/stw887.

(12) Wakelam, V.; Loison, J. C.; Herbst, E.; Pavone, B.; Bergeat, A.; Béroff, K.; Chabot, M.; Faure, A.; Galli, D.; Geppert, W. D.; et al. The 2014 KIDA Network for Interstellar Chemistry. *Astrophys. J., Suppl. Ser.* **2015**, *217*, 20. DOI: 10.1088/0067-0049/217/2/20.

(13) Willis, R. H. J.; Speak, T. H.; Byrne, A. N.; Shingledecker, C. N.; Cooke, I. R. The Impact of Hydrogen Atom Tunneling on Aromatic Chemistry in TMC-1. *Astrophys. J.,* **2026**, *1005*, 241.

(14) Loison, J.-C.; Jacovella, U.; Rossi, C.; Rasmussen, A. P.; Thissen, R.; Pattathadathil, N.; Alcaraz, C.; Wakelam, V. Formation of Benzene and its Derivatives in Dense Molecular Clouds. *Astron. Astrophys.* **2026**, *709*, 10.1051/0004-6361/202659618.

(15) Becker, K. H.; Donner, B.; Dinis, C. M. F.; Geiger, H.; Schmidt, F.; Wiesen, P. Kinetics of the $C_2(a^3\Pi_u)$ Radical Reacting with Selected Molecules and Atoms. *Z. Phys. Chem.* **2000**, *214*, 503 - 517.

(16) Berman, M. R.; Fleming, J. W.; Harvey, A. B.; Lin, M. C. Temperature Dependence of the Reactions of CH Radicals with Unsaturated Hydrocarbons. *Chem. Phys.* **1982**, *73*, 27-33.

(17) Hamon, S.; Le Picard, S. D.; Canosa, A.; Rowe, B. R.; Smith, I. W. M. Low Temperature Measurements of the Rate of Association to Benzene Dimers in Helium. *J. Chem. Phys.* **2000**, *112*, 4506-4516.

(18) Goulay, F.; Leone, S. R. Low-Temperature Rate Coefficients for the Reaction of Ethynyl Radical ($C_2H$) with Benzene. *J. Phys. Chem. A* **2006**, *110*, 1875-1880, Article. DOI: 10.1021/jp055637p.

(19) Manion, J. A.; Huie, R. E.; Levin, R. D.; Burgess Jr., D. R.; Orkin, V. L.; Tsang, W.; McGivern, W. S.; Hudgens, J. W.; Knyazev, V. D.; Atkinson, D. B.; et al. *NIST Chemical Kinetics Database,* NIST Standard Reference Database 17, Version 7.0 (Web Version), Release 1.6.8, Data version 2015.09, National Institute of Standards and Technology, Gaithersburg, Maryland, 20899-8320. Web address: https://kinetics.nist.gov/

(20) Haider, N.; Husain, D. Kinetic Investigation of the Collisional Behavior of Ground State Atomic Carbon, $C(2p^2(^3P_j))$, with Halogenated Olefins and Aromatic Compounds Studied by Time-Resolved Atomic Resonance Absorption Spectroscopy in the Vacuum Ultra-Violet. *Int. J. Chem. Kin.* **1993**, *25*, 423-435.

(21) Wenzel, G.; Speak, T. H.; Changala, P. B.; Willis, R. H. J.; Burkhardt, A. M.; Zhang, S.; Bergin, E. A.; Byrne, A. N.; Charnley, S. B.; Fried, Z. T. P.; et al. Detections of Interstellar Aromatic Nitriles 2-cyanopyrene and 4-cyanopyrene in TMC-1. *Nat. Astron.* **2025**, *9*, 262-270.

(22) Kaiser, R. I. Crossed Beams Reaction of Atomic Carbon, $C(^3P_j)$, with $d_6$-Benzene, $C_6D_6(X^1A_{1g})$: Observation of the Per-Deutero-1,2-Didehydro-Cycloheptatrienyl Radical, $C_7D_5(X^2B_2)$. *J. Chem. Phys.* **1999**, *110*, 6091-6094.

(23) Hahndorf, I.; Lee, Y. T.; Kaiser, R. I.; Vereecken, L.; Peeters, J.; Bettinger, H. F.; Schreiner, P. R.; Schleyer, P. V. R.; Allen, W. D.; Schaefer Iii, H. F. A Combined Crossed-Beam, Ab Initio, and Rice-Ramsperger-Kassel-Marcus Investigation of the Reaction of Carbon Atoms C($^3P_j$) with Benzene, $C_6H_6$($X^1A_{1g}$) and D6-Benzene, $C_6D_6$($X^1A_{1g}$). *J. Chem. Phys.* **2002**, *116*, 3248-3262.
(24) Krasnokutski, S. A.; Huisken, F. Ultra-Low-Temperature Reactions of C($^3P_0$) Atoms with Benzene Molecules in Helium Droplets. *J. Chem. Phys.* **2014**, *141*, 214306. DOI: doi:http://dx.doi.org/10.1063/1.4902369.
(25) Bergeat, A.; Loison, J.-C. Reaction of Carbon Atoms, C ($2p^2$, $^3P$) with $C_2H_2$, $C_2H_4$ and $C_6H_6$: Overall Rate Constant and Relative Atomic Hydrogen Production. *Phys. Chem. Chem. Phys.* **2001**, *3*, 2038-2042. DOI: 10.1039/b100656h.
(26) da Silva, G. Reaction of Benzene with Atomic Carbon: Pathways to Fulvenallene and the Fulvenallenyl Radical in Extraterrestrial Atmospheres and the Interstellar Medium. *J. Phys. Chem. A* **2014**, *118*, 3967-3972. DOI: 10.1021/jp503431a.
(27) Izadi, M. E.; Bal, K. M.; Maghari, A.; Neyts, E. C. Reaction Mechanisms of C($^3P_j$) and $C^+$($^2P_j$) with Benzene in the Interstellar Medium from Quantum Mechanical Molecular Dynamics Simulations. *Phys. Chem. Chem. Phys.* **2021**, *23*, 4205-4216, 10.1039/D0CP04542J. DOI: 10.1039/D0CP04542J.
(28) Daugey, N.; Caubet, P.; Bergeat, A.; Costes, M.; Hickson, K. M. Reaction Kinetics to Low Temperatures. Dicarbon + Acetylene, Methylacetylene, Allene and Propene from $77 \leq T \leq 296$ K. *Phys. Chem. Chem. Phys.* **2008**, *10*, 729-737. DOI: 10.1039/b710796j.
(29) Daugey, N.; Caubet, P.; Retail, B.; Costes, M.; Bergeat, A.; Dorthe, G. Kinetic Measurements on Methylidyne Radical Reactions with Several Hydrocarbons at Low Temperatures. *Phys. Chem. Chem. Phys.* **2005**, *7*, 2921-2927. DOI: 10.1039/b506096f.
(30) Hickson, K. M.; Loison, J.-C. Kinetic Study of the Reactions of Ground State Atomic Carbon and Oxygen with Nitrogen Dioxide over the 50–296 K Temperature Range. *J. Phys. Chem. A* **2024**, *128*, 10598-10608. DOI: 10.1021/acs.jpca.4c06193.
(31) Shannon, R. J.; Cossou, C.; Loison, J.-C.; Caubet, P.; Balucani, N.; Seakins, P. W.; Wakelam, V.; Hickson, K. M. The Fast C($^3P$) + $CH_3OH$ Reaction as an Efficient Loss Process for Gas-Phase Interstellar Methanol. *RSC Adv.* **2014**, *4*, 26342-26353. DOI: 10.1039/c4ra03036b.
(32) Hickson, K. M.; Loison, J.-C.; Wakelam, V. Kinetic Study of the Gas-Phase Reaction between Atomic Carbon and Acetone: Low-Temperature Rate Constants and Hydrogen Atom Product

Yields. *ACS Earth Space Chem.* **2023**, *7*, 2091-2104. DOI: 10.1021/acsearthspacechem.3c00193.

(33) Ernstberger, B.; Krause, H.; Neusser, H. J. Metastable Decay and Binding Energies of Van Der Waals Cluster Ions. *Z. Phys. D: At., Mol. Clusters* **1991**, *20*, 189-192. DOI: 10.1007/BF01543970.

(34) Weber, T.; Smith, A. M.; Riedle, E.; Neusser, H. J.; Schlag, E. W. High-Resolution UV Spectrum of the Benzene—$N_2$ Van Der Waals Complex. *Chem. Phys. Lett.* **1990**, *175*, 79-83. DOI: https://doi.org/10.1016/0009-2614(90)85521-D.

(35) Etzkorn, T.; Klotz, B.; Sørensen, S.; Patroescu, I. V.; Barnes, I.; Becker, K. H.; Platt, U. Gas-Phase Absorption Cross Sections of 24 Monocyclic Aromatic Hydrocarbons in the UV and IR Spectral Ranges. *Atmos. Environ.* **1999**, *33*, 525-540. DOI: https://doi.org/10.1016/S1352-2310(98)00289-1.

(36) Davico, G. E.; Bierbaum, V. M.; DePuy, C. H.; Ellison, G. B.; Squires, R. R. The C-H Bond Energy of Benzene. *J. Am. Chem. Soc.* **1995**, *9*, 2590-2599.

(37) Hickson, K. M.; Loison, J.-C.; Lique, F.; Kłos, J. An Experimental and Theoretical Investigation of the $C(^1D) + N_2 \rightarrow C(^3P) + N_2$ Quenching Reaction at Low Temperature. *J. Phys. Chem. A* **2016**, *120*, 2504-2513. DOI: 10.1021/acs.jpca.6b00480.

(38) Bettinger, H. F.; Schleyer, P. V. R.; Schaefer Iii, H. F.; Schreiner, P. R.; Kaiser, R. I.; Lee, Y. T. Reaction of Benzene with a Ground State Carbon Atom, $C(^3P_j)$. *J. Chem. Phys.* **2000**, *113*, 4250-4264.

(39) Millar, T. J.; Walsh, C.; Van de Sande, M.; Markwick, A. J. The UMIST Database for Astrochemistry 2022. *Astron. Astrophys.* **2024**, *682*, 10.1051/0004-6361/202346908.

(40) Wakelam, V.; Gratier, P.; Loison, J. C.; Hickson, K. M.; Penguen, J.; Mechineau, A. The 2024 KIDA Network for Interstellar Chemistry. *Astron. Astrophys.* **2024**, *689*, 10.1051/0004-6361/202450606.

(41) Coutens, A.; Loison, J. C.; Boulanger, A.; Caux, E.; Müller, H. S. P.; Wakelam, V.; Manigand, S.; Jørgensen, J. K. The Alma-PILS Survey: First Tentative Detection of 3-Hydroxypropenal (HOCHCHCHO) in the Interstellar Medium and Chemical Modeling of the $C_3H_4O_2$ Isomers. *Astron. Astrophys.* **2022**, *660*, 10.1051/0004-6361/202243038.

(42) Manigand, S.; Coutens, A.; Loison, J. C.; Wakelam, V.; Calcutt, H.; Müller, H. S. P.; Jørgensen, J. K.; Taquet, V.; Wampfler, S. F.; Bourke, T. L.; et al. The Alma-PILS Survey: First

Detection of the Unsaturated 3-Carbon Molecules Propenal ($C_2H_3CHO$) and Propylene ($C_3H_6$) Towards IRAS 16293–2422 B. *Astron. Astrophys.* **2021**, *645*, 10.1051/0004-6361/202038113.
(43) Hickson, K. M.; Loison, J.-C.; Wakelam, V. Kinetic Study of the Gas-Phase C($^3$P) + $CH_3CN$ Reaction at Low Temperatures: Rate Constants, H-atom Product Yields, and Astrochemical Implications. *ACS Earth Space Chem.* **2021**, *5*, 824-833. DOI: 10.1021/acsearthspacechem.0c00347.
(44) Hickson, K. M.; Loison, J.-C.; Wakelam, V. A Low-Temperature Kinetic Study of the C($^3$P) + $CH_3OCH_3$ Reaction: Rate Constants, H Atom Product Yields, and Astrochemical Implications. *ACS Earth Space Chem.* **2024**, *8*, 1087-1100. DOI: 10.1021/acsearthspacechem.4c00014.
(45) Hincelin, U.; Wakelam, V.; Hersant, F.; Guilloteau, S.; Loison, J. C.; Honvault, P.; Troe, J. Oxygen Depletion in Dense Molecular Clouds: A Clue to a Low $O_2$ abundance? *Astron. Astrophys.* **2011**, *530*, A61. DOI: 10.1051/0004-6361/201016328.
(46) Byrne, A. N.; Shingledecker, C. N.; Bergin, E. A.; Holdren, M. S.; Wenzel, G.; Xue, C.; Van Voorhis, T.; McGuire, B. A. Modeling the Effect of C/O Ratio on Complex Carbon Chemistry in Cold Molecular Clouds. *Astrophys. J.* **2026**, *998*, 95. DOI: 10.3847/1538-4357/ae2fed.
(47) Byrne, A. N.; Xue, C.; Van Voorhis, T.; McGuire, B. A. Sensitivity Analysis of Aromatic Chemistry to Gas-Phase Kinetics in a Dark Molecular Cloud Model. *Phys. Chem. Chem. Phys.* **2024**, *26*, 26734-26747, 10.1039/D4CP03229B. DOI: 10.1039/D4CP03229B.
(48) Mallo, M.; Agúndez, M.; Cabezas, C.; Roncero, O.; Cernicharo, J.; Molpeceres, G. Ion-Molecule Routes Towards Cycles in TMC-1. *Astron. Astrophys.* **2025**, *704*, 10.1051/0004-6361/202557647.
(49) Hincelin, U.; Commerçon, B.; Wakelam, V.; Hersant, F.; Guilloteau, S.; Herbst, E. Chemical and Physical Characterization of Collapsing Low-Mass Prestellar Dense Cores. *Astrophys. J.* **2016**, *822*, 12. DOI: 10.3847/0004-637X/822/1/12.
(50) Wakelam, V.; Ruaud, M.; Gratier, P.; Bonnell, I. A. Influence of Galactic Arm Scale Dynamics on the Molecular Composition of the Cold and Dense ISM – II. Molecular Oxygen Abundance. *Mon. Not. R. Astron. Soc.* **2019**, *486*, 4198-4202. DOI: 10.1093/mnras/stz1122.
(51) Wakelam, V.; Dartois, E.; Chabot, M.; Spezzano, S.; Navarro-Almaida, D.; Loison, J. C.; Fuente, A. Efficiency of Non-Thermal Desorptions in Cold-Core Conditions. *Astron. Astrophys.* **2021**, *652*, 10.1051/0004-6361/202039855.
(52) Vuitton, V.; Yelle, R. V.; Lavvas, P.; Klippenstein, S. J. Rapid Association Reactions at Low Pressure: Impact on the Formation of Hydrocarbons on Titan. *Astrophys. J.* **2012**, *744*, 11.

(53) Anicich, V.G., 2003. An index of the literature for bimolecular gas phase cation-molecule reaction kinetics. JPL Publication 03-19, 1–1194.

(54) Cernicharo, J.; Agúndez, M.; Cabezas, C.; Tercero, B.; Marcelino, N.; Pardo, J. R.; de Vicente, P. Pure Hydrocarbon Cycles in TMC-1: Discovery of Ethynyl Cyclopropenylidene, Cyclopentadiene, and Indene. *Astron. Astrophys.* **2021**, *649*, 10.1051/0004-6361/202141156.

(55) Cernicharo, J.; Agúndez, M.; Kaiser, R. I.; Cabezas, C.; Tercero, B.; Marcelino, N.; Pardo, J. R.; de Vicente, P. Discovery of Benzyne, o-$C_6H_4$, in TMC-1 with the Quijote Line Survey. *Astron. Astrophys.* **2021**, *652*, 10.1051/0004-6361/202141660.

(56) Cernicharo, J.; Fuentetaja, R.; Agúndez, M.; Kaiser, R. I.; Cabezas, C.; Marcelino, N.; Tercero, B.;Pardo, J. R.; de Vicente, P. Discovery of Fulvenallene in TMC-1 with the QUIJOTE Line Survey. *Astron. Astrophys.* **2022**, *663*, L9.

(57) Cernicharo, J.; Agúndez, M.; Kaiser, R. I.; Cabezas, C.; Tercero, B.; Marcelino, N.; Pardo, J. R.; de Vicente, P. Discovery of Two Isomers of Ethynyl Cyclopentadiene in TMC-1: Abundances of CCH and CN Derivatives of Hydrocarbon Cycles. *Astron. Astrophys.* **2021**, *655*, L1.

(58) Loru, D.; Cabezas, C.; Cernicharo, J.; Schnell, M.; Steber, A. L. Detection of Ethynylbenzene in TMC-1 and the Interstellar Search for 1,2-Diethynylbenzene. *Astron. Astrophys.* **2023**, *677*, 10.1051/0004-6361/202347023.

(59) Smith, R. D.; Futrell, J. H. Reactions of Thermal Energy ($^2P$) $C^+$ Ions with Several Molecules. *Int. J. Mass Spectrom. Ion Phys.* **1978**, *26*, 111-113.

(60) Hamberg, M.; Vigren, E.; Thomas, R. D.; Zhaunerchyk, V.; Zhang, M.; Trippel, S.; Kaminska, M.; Kashperka, I.; af Ugglas, M.; Källberg, A.; et al. Experimental Studies of the Dissociative Recombination Processes for the $C_6D_6^+$ and $C_6D_7^+$ Ions. EAS Publications Series **2011**, *46*, 241-249.

(61) Jacovella, U.; Loison, J.-C.; Rossi, C.; Rasmussen, A. P.; Thissen, R.; Pattathadathil, N.; Alcaraz, C. $C_5H_6$ Formation in Dense Molecular Clouds. *Astron. Astrophys.* **2026**, *708*, 10.1051/0004-6361/202658985.

(62) He, C.; Thomas, A. M.; Galimova, G. R.; Morozov, A. N.; Mebel, A. M.; Kaiser, R. I. Gas-Phase Formation of Fulvenallene ($C_7H_6$) Via the Jahn–Teller Distorted Tropyl ($C_7H_7$) Radical Intermediate under Single-Collision Conditions. *J. Am. Chem. Soc.* **2020**, *142*, 3205-3213. DOI: 10.1021/jacs.9b13269.

(63) Harding, L. B.; Klippenstein, S. J.; Georgievskii, Y. On the Combination Reactions of Hydrogen Atoms with Resonance-Stabilized Hydrocarbon Radicals. *J. Phys. Chem. A* **2007**, *111*, 3789 - 3801.